\documentclass[lettersize,journal]{IEEEtran}
\IEEEoverridecommandlockouts

\usepackage{url}
\usepackage{cite}
\usepackage{makecell}
\usepackage{subfigure}
\usepackage{subfig}
\usepackage{amsmath,amssymb,amsfonts}
\usepackage{amsthm}
\usepackage{algorithmic}
\usepackage{algorithm}

\usepackage{graphicx}
\usepackage{textcomp}
\usepackage{booktabs}
\usepackage{bm}
\usepackage{float}  
\usepackage{dblfloatfix} 
\usepackage{pifont}
\usepackage{comment}
\usepackage{cellspace}

\usepackage{wasysym,makecell}

\usepackage{multirow}
\usepackage[normalem]{ulem}
\useunder{\uline}{\ul}{}
\usepackage{xcolor}

\def\BibTeX{{\rm B\kern-.05em{\sc i\kern-.025em b}\kern-.08em
    T\kern-.1667em\lower.7ex\hbox{E}\kern-.125emX}}
\allowdisplaybreaks

\begin{document}

\title{
Communications-Incentivized Collaborative Reasoning in NetGPT through Agentic Reinforcement Learning
}
\author{Xiaoxue Yu, Rongpeng Li, Zhifeng Zhao, and Honggang Zhang 
\vspace{-.5cm}
\\

\thanks{This work has been submitted to the IEEE for possible publication. Copyright may be transferred without notice, after which this version may no longer be accessible.}
}
\maketitle

\begin{abstract}
The evolution of next-Generation (xG) wireless networks marks a paradigm shift from connectivity-centric architectures to Artificial Intelligence (AI)-native designs that tightly integrate data, computing, and communication. Yet existing AI deployments in communication systems remain largely siloed, offering isolated optimizations without intrinsic adaptability, dynamic task delegation, or multi-agent collaboration. In this work, we propose a unified agentic NetGPT framework for AI-native xG networks, wherein a NetGPT core can either perform autonomous reasoning or delegate sub-tasks to domain-specialized agents via agentic communication. The framework establishes clear modular responsibilities and interoperable workflows, enabling scalable, distributed intelligence across the network. To support continual refinement of collaborative reasoning strategies, the framework is further enhanced through Agentic reinforcement learning under partially observable conditions and stochastic external states. The training pipeline incorporates masked loss against external agent uncertainty, entropy-guided exploration, and multi-objective rewards that jointly capture task quality, coordination efficiency, and resource constraints. Through this process, NetGPT learns when and how to collaborate, effectively balancing internal reasoning with agent invocation. Overall, this work provides a foundational architecture and training methodology for self-evolving, AI-native xG networks capable of autonomous sensing, reasoning, and action in complex communication environments. 
\end{abstract}

\section{Introduction}\label{sec1}
The evolution of next-generation (xG) wireless networks marks a paradigm shift from connectivity-centric architectures to Artificial Intelligence (AI)-native designs that tightly integrate data, computing, and communication. Recent efforts such as NetGPT \cite{NetGPT} highlight the potential of synergizing cloud and edge Large Language Models (LLMs) to enable personalized, context-aware generative services and to support intelligent network management and resource orchestration \cite{netgpt_resource_management}. 
Concurrent developments also start to introduce large vision and multi-modal models into communication systems \cite{FoundationModel4Net}. 
In parallel, domain-specialized agents with localized perception and decision-making capabilities are emerging as key execution units \cite{AaaS}. Their effectiveness, however, relies not only on their internal reasoning but also on efficient communication and coordination across diverse network environments.

Agentic communications is therefore becoming essential. On the content side, agents may exchange structured semantic representations such as embeddings, task instructions, sub-agent invocation requests, or tool-call metadata, rather than merely raw bit streams. On the protocol side, emerging interoperability standards, including the Agent-to-Agent Protocol (A2A), the Agent Communication Protocol (ACP), and the Agent Network Protocol (ANP) \cite{AgentCommSurvey}, are being developed to support large-scale agent discovery, messaging, and orchestration. Yet, current agentic systems still rely on static, prompt-driven roles whose coordination degrades over time \cite{llm_multi_agents_survey}. Collaborative reasoning must contend with unstable or noisy signals arising from multi-agent interactions, evolving contexts, and resource or communication constraints. These characteristics introduce substantial challenges associated with non-deterministic state transitions and with determining when an agent should rely on internal reasoning versus seeking external collaboration.
To tackle these challenges and achieve adaptive, resilient, and scalable collaborative reasoning, continual learning will become indispensable. 

In general, continual learning naturally falls within the scope of Reinforcement Learning (RL). Conventional reinforcement tuning methods, such as PPO used in RL from Human Feedback (RLHF) \cite{RLHF} and GRPO \cite{DeepseekR1} used in RL with Verifiable Rewards (RLVR), have achieved remarkable improvements in single-agent reasoning. 
However, these methods typically assume deterministic state transitions, where the state corresponds to the accumulated token sequence and the next state is fully determined by concatenation. In contrast, agents engaged in communications-driven collaborative reasoning are no longer passive generators; rather, they are embedded within sequential decision-making loops \cite{AgenticRL_survey}. Moreover, the agent's action extends far beyond text generation to include structured actions such as tool invocation, environment manipulation, and task delegation, while rewards may be sparse or delayed. This shift motivates the adoption of Agentic Reinforcement Learning (Agentic RL), which operates under partial observability and stochastic dynamics, where external observations, communication content, and environmental feedback continuously reshape the state.

\begin{figure*}[!ht]
	\centering 
	\includegraphics[scale = 0.74]{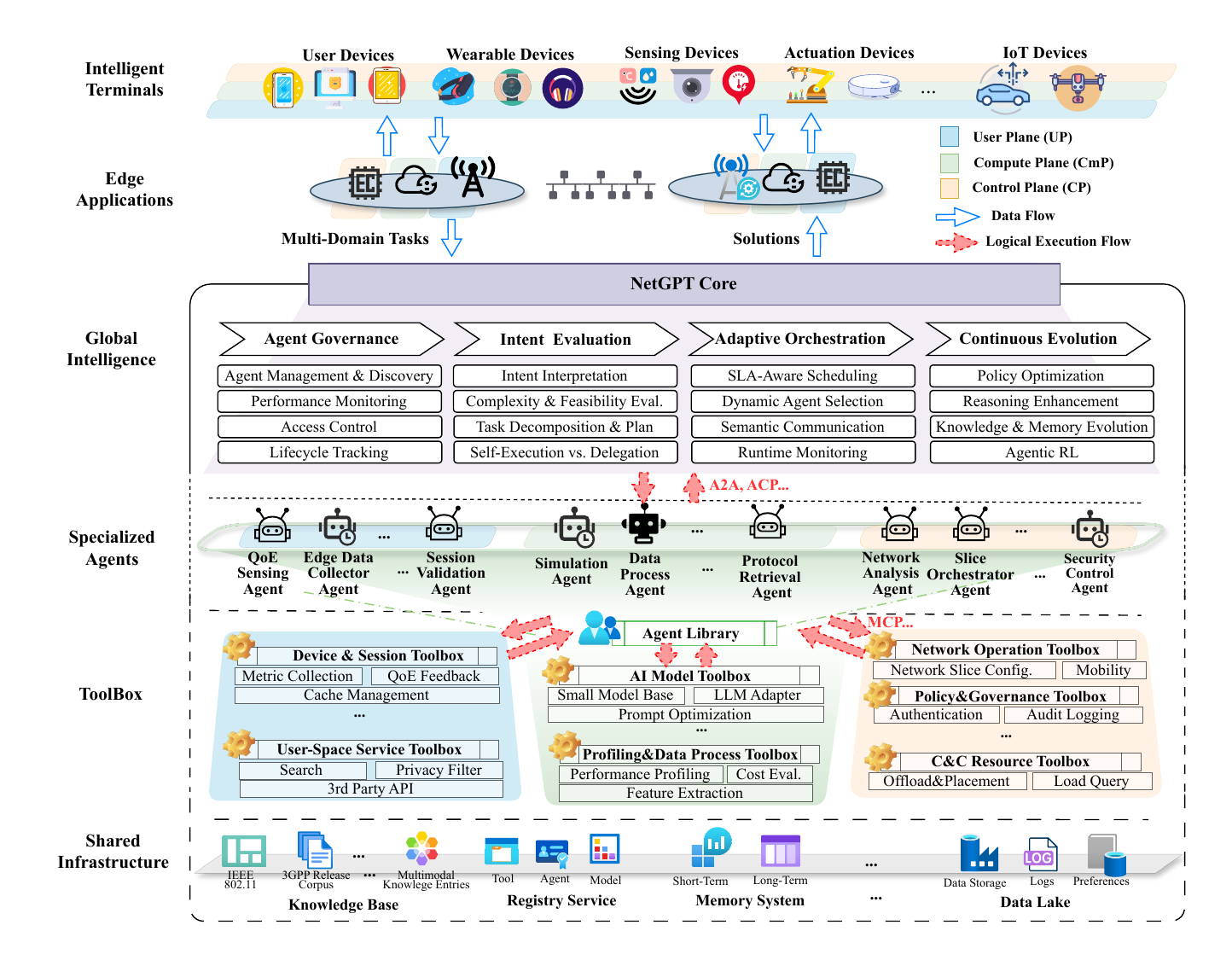}
	\caption{Communications-incentivized collaborative reasoning in NetGPT.}
	\label{fig1:framework}
	\vspace{-1.2em}
\end{figure*} 

In this work, we propose a unified agentic NetGPT framework for AI-native networks, in which a NetGPT core can either perform independent reasoning or delegate subtasks to domain-specialized agents through agentic communication. 
Unlike prior work on tool-augmented reasoning \cite{llm_with_tool_survey}, this framework delineates the responsibilities of the NetGPT core and the domain-specific agents and their toolboxes, elevating collaboration from simple tool calling to agent-driven collaborative problem solving within a distributed xG intelligence fabric. 
Building on this architecture, we integrate agentic RL to enhance NetGPT's collaborative reasoning. To further improve the training stability and effectiveness in stochastic and partially observable environments, we make substantial contributions to agentic RL by incorporating dynamic masked loss, entropy-guided exploration, and multi-objective rewards. 
Together, the framework and mechanism provide both an architectural foundation and a practical agentic RL pipeline that advances agentic intelligence in xG networks.

\section{Communications-based Collaborative Reasoning Framework in NetGPT} 

\subsection{Framework and Components of NetGPT}
\begin{figure*}[!ht]
	\centering 
	\hspace*{-0.2cm}
	\includegraphics[scale = 0.5]{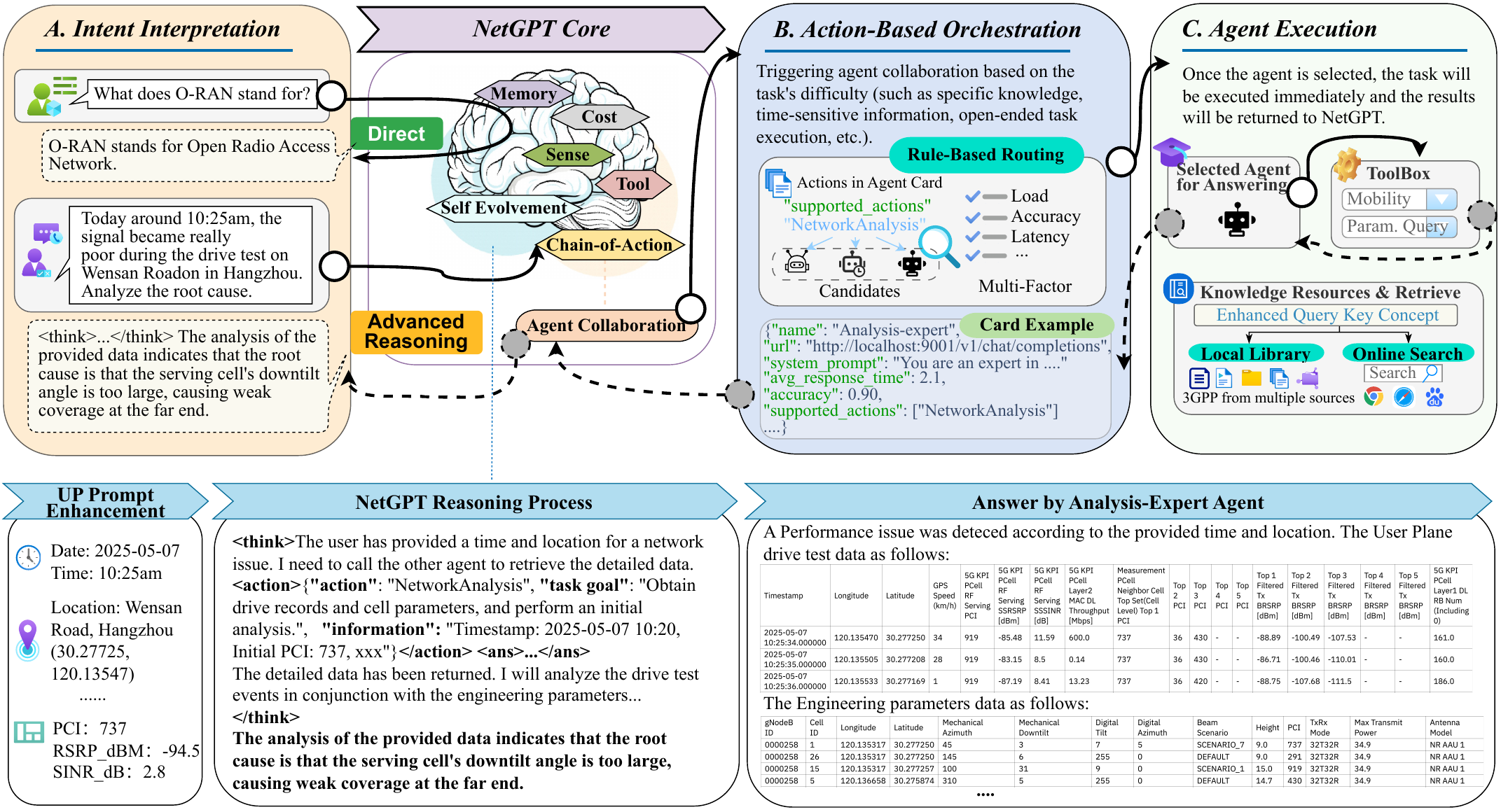}
	\caption{(\textbf{Top}): NetGPT core-based collaborative reasoning workflow from intent interpretation, action-based orchestration, and execution. (\textbf{Bottom}): Collaborative reasoning example for network root cause analysis.}
	\vspace{-.4cm}
	\label{fig2:operation details}
\end{figure*}

As illustrated in Fig.~\ref{fig1:framework}, NetGPT operates across multiple layers \cite{NetGPT}, with the NetGPT core serving as the primary service endpoint for terminals and edge applications, handling both internal and external reasoning requests. 
Before forwarding requests to the NetGPT core, as shown in Fig. \ref{fig2:operation details}, edge applications may run lightweight clients that package, authenticate, and enrich requests with localized context, thereby reducing local processing overhead while ensuring sufficient information completeness. 

Beyond the NetGPT core, a wide range of domain-specialized agents and toolboxes are distributed across the network. These agents primarily reside in the compute plane and are organized within an agent library that the NetGPT core can instantiate, coordinate, and retire on demand. As shown in Fig.~\ref{fig1:framework}, each network plane hosts agents tailored to its operational responsibilities while remaining discoverable and callable by the NetGPT core. For instance, user-plane agents process real-time data flows between end devices and network functions; compute-plane agents execute analytical, predictive, or simulation-intensive workloads; and control-plane agents implement configuration management and closed-loop control. These agents provide contextual awareness, communication interfaces, and decision autonomy within each network plane. Meanwhile, standard agent-to-agent communication protocols (e.g., A2A, ACP, and ANP \cite{AgentCommSurvey}) further enable decentralized yet coherent operation across planes. 

Complementing these specialized agents, toolboxes encapsulate reusable functional modules and constitute a lightweight and replaceable operational foundation.
Common toolboxes include device and session utilities in the user plane, analytical and AI-model utilities in the compute plane, and operational and policy mechanisms in the control plane, etc. Accessed through agent-environment protocols such as the Model Context Protocol (MCP), these toolboxes integrate learning-based intelligence with traditional network operations.

Lastly, the shared infrastructure layer offers common services such as knowledge bases, registry functions, and data lakes, enabling global knowledge access and persistent memory for the NetGPT core, specialized agents, and toolboxes across all planes.

\subsection{Key Functionalities for Collaborative Reasoning}
As illustrated in Fig. \ref{fig2:operation details}, when a contextually-enhanced request arrives at the NetGPT core, it first interprets and enhances the user intent. 
Depending on the complexity and difficulty of the task, the core may answer directly or proceed to determine the required execution paths and the decomposed sub-tasks pertaining to registered action types. By querying agent cards that declare support for the specific action capability, NetGPT selects an agent, orchestrates the execution through structured communication protocols, monitors progress, and integrates results into a coherent final response. 
Therefore, fulfilling collaborative reasoning encompasses several essential procedures, spanning from agent governance, intent evaluation, adaptive orchestration, and agent execution, as well as continual evolution.

\subsubsection{Agent Governance}
Instead of binding the registry to a single interoperability protocol, NetGPT employs a protocol-agnostic agent registry. Differences among protocols such as A2A, ACP, and ANP \cite{AgentCommSurvey} can be absorbed through adapters that map agent identities, capability descriptions, invocation endpoints, and other metadata into a unified agent card. 
Each agent card maintains functional capabilities and xG-related metadata (e.g., throughput, latency, load, and cost), enabling network-aware agent selection under the same action type while supporting agent management, access control, performance monitoring, and lifecycle tracking.

\subsubsection{Intent Evaluation}
Rather than functioning as a simple routing component, the NetGPT core evaluates task intent and complexity to determine whether to execute the task locally or decompose it into sub-tasks for registered agents. This prevents unnecessary reasoning depth and avoids additional latency when local execution suffices.

\subsubsection{Adaptive Orchestration}
Once intent evaluation determines the sub-task decomposition, NetGPT queries the registered agent cards to identify all candidate agents supporting the required action type. It then selects the most appropriate agent through a routing scheme that can incorporate network conditions, task requirements, and other contextual signals. 
Among several routing mechanisms \cite{AgentCommSurvey}, rule-based routing applies deterministic matching logic, offering transparency and reliability for stable or safety-critical applications. Machine learning-based routing uses data-driven classifiers trained on labeled intent-routing pairs, improving generalization but requiring continual retraining. LLM-based routing leverages contextual reasoning for the highest flexibility, albeit at the cost of increased latency and inference overhead. 
Through this action-oriented (rather than identity-based) discovery and routing process, NetGPT achieves natural extensibility: any new agent that declares the required action type can participate in collaborative reasoning without necessitating further model retraining, enabling scalable task delegation that adapts to evolving service requirements, resource availability, and system dynamics. 

\subsubsection{Continuous Evolution}
NetGPT supports continual improvement of its decision-making and coordination strategies. 
Through agentic RL, it evaluates performance metrics and feedback signals to refine its collaborative reasoning behavior, learning when to invoke external agents and how to balance accuracy, latency, and resource cost, thereby establishing a sustained improvement loop beyond fixed prompt-engineered systems.

\begin{figure}[t]
	\centering 
	\includegraphics[scale = 0.55]{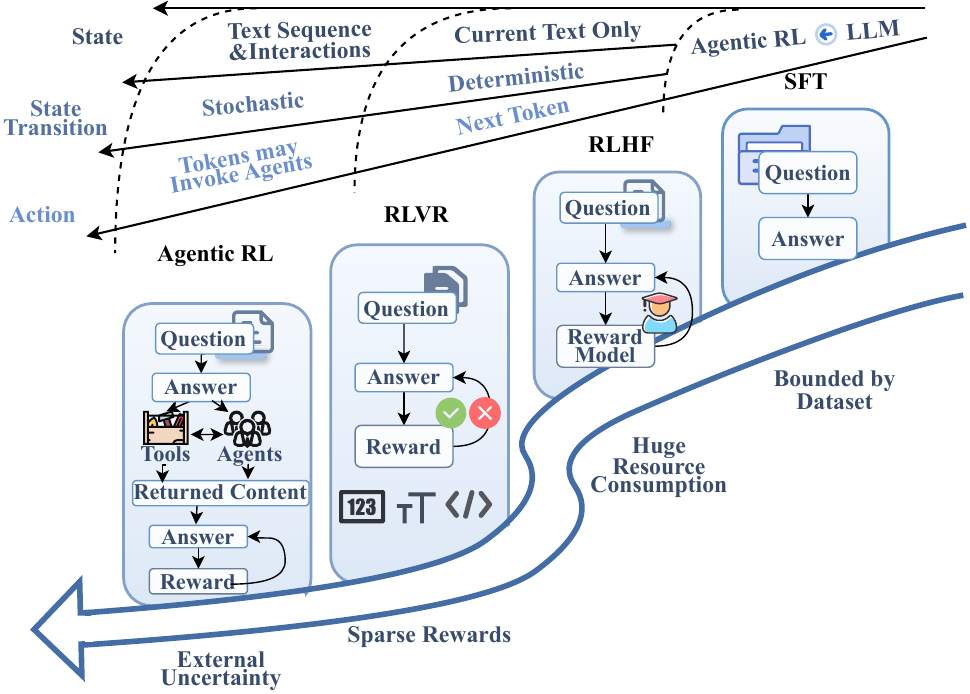}
	\caption{Development of training phase for LLM agents.}
\label{fig3:development_phase}
\end{figure} 

\begin{figure*}[t]
	\centering 
	\hspace*{0.3cm}
	\includegraphics[scale = 0.43]{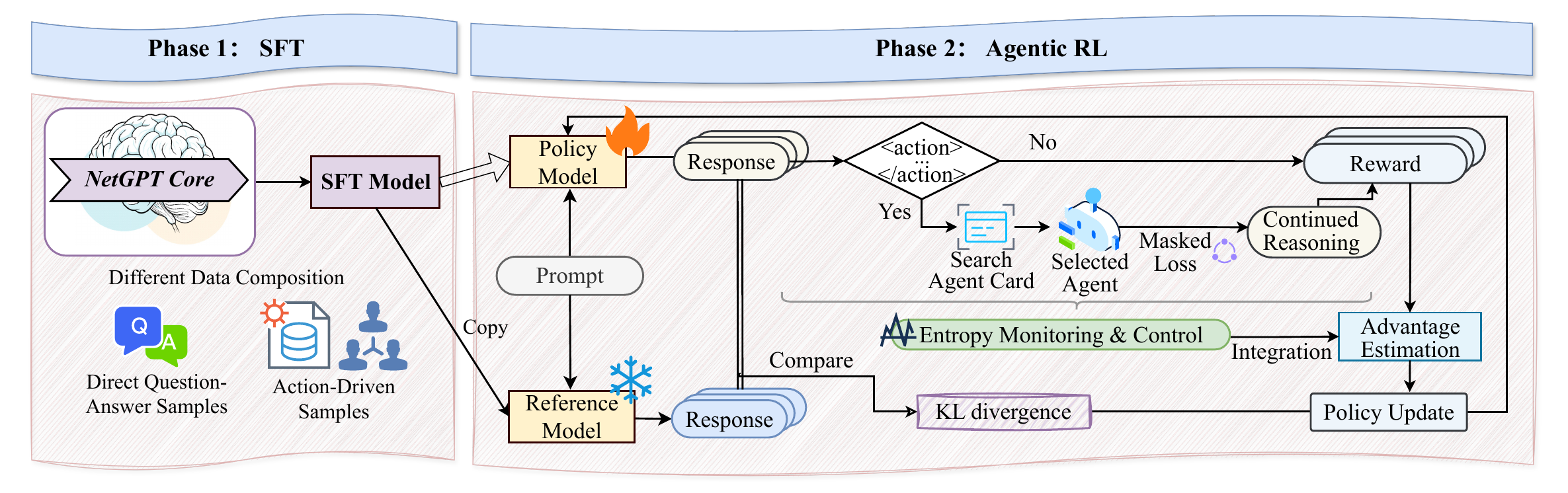}
	\caption{Training pipeline for NetGPT.}
\label{fig4:training_pipeline}
\end{figure*} 

\section{Continual Evolvement for NetGPT via Agentic RL}
\label{sec:training}
\subsection{From SFT to Agentic RL}
Before delving into the specifics of agentic RL for NetGPT, it is instructive to review the evolution of training paradigms for LLM-based agents, as illustrated in Fig.~\ref{fig3:development_phase}. The emergence of agents capable of sophisticated reasoning has been closely linked to the maturation of RL-inspired training methods that have progressively expanded flexibility, exploration capacity, and decision-making capabilities.

Initially, Supervised Fine-Tuning (SFT) endows models with instruction-following abilities using domain-specific, labeled prompt-response pairs under a supervised loss objective. However, reasoning remained largely limited to patterns present in the training data and lacked adaptability beyond dataset coverage. Moving beyond pure imitation and token-likelihood optimization, RL-based approaches introduce exploration and outcome-driven optimization, enabling alignment with more complex reasoning tasks. For instance, RLHF optimizes an LLM as an RL policy using human preference rewards that target long-term outcomes \cite{RLHF}, whereas RLVR removes the dependence on a separate reward model by using objective and verifiable signals, such as correctness checks, execution results, or symbolic validation, as direct reward sources \cite{DeepseekR1}. Nevertheless, both RLHF and RLVR remain constrained to prompt-response interactions and suffer from sparse terminal rewards, and cannot support the effectiveness of multi-step reasoning and interactive decision-making required by NetGPT. In the collaborative reasoning process, NetGPT needs to choose actions and interact with other agents across reasoning-action loops beyond text-generation, while maintaining coherent long-horizon reasoning.

The aforementioned limitations motivate the shift toward agentic RL, where tool use and intermediate verification become part of the trajectory and are optimized jointly with the model's reasoning behavior. 
For example, ARTIST \cite{ARTIST} enables models not only to produce solutions but also to decide when and how to invoke external tools such as code interpreters, APIs, or search engines during multi-turn reasoning via outcome-based RL (e.g., GRPO). Well-balanced entropy-aware exploration allows adaptive branching at uncertain steps and assigns credit across shared and branched reasoning paths \cite{ARPO}, eliminating possible over-branching and unstable gradients. These advances make agentic RL a promising solution for training NetGPT. 

\subsection{Training Pipeline}
\label{sec:training_pipeline}
As shown in Fig.~\ref{fig4:training_pipeline}, the training of the NetGPT core follows a two-phase pipeline that combines SFT and agentic RL optimization. 

In the first \emph{supervised warm-up} phase, the objective is to enable the NetGPT core to acquire the foundational behavioral and structural, and instruction-following capabilities required for agentic reasoning. This stage relies on a deliberately balanced data composition that exposes the model to multiple levels of interaction complexity, including:
\begin{itemize}
	\item Direct Question-Answer Samples: 
	These samples demonstrate how to produce concise and accurate responses to straightforward query-answer pairs. 
\item Action-Driven Samples: 
These samples contain dialogues annotated with explicit action invocations via \texttt{<action>} and \texttt{</action>}, training the model to select predefined ``\texttt{supported\_actions}" types and invoke agents rather than answering directly. 
\end{itemize}

Through this process, the model learns to align user requests with predefined action types and to provide necessary task goals and contextual information for downstream agents, as shown in Fig. \ref{fig2:operation details}. 
Subsequently, the NetGPT core is further optimized through agentic RL to enhance decision-making, collaboration efficiency, and reasoning adaptability. During this stage, the system can interact with either real or simulated xG network environments and learns from the predefined reward signals, enabling it to determine better when to perform self-execution and when to delegate tasks to external agents. 
To further improve learning stability and overall training efficiency, several additional components are incorporated into the optimization pipeline.  

\subsubsection{Routing}
When an invoking action is triggered, a rule-based routing policy, as shown in Fig. \ref{fig2:operation details}, is adopted due to its scalability, by selecting candidate agents based on the ``\texttt{supported\_actions}" metadata recorded in their agent cards. As training progresses, the routing criteria can gradually adapt to different task profiles by using a weighted combination of metrics such as current load, historical accuracy, and average response latency. This adaptive weighting allows the system to prioritize factors most relevant to real-time control scenarios, leading to deterministic selections that balance both performance and efficiency.

\subsubsection{Masked Loss for Agent Responses}
The output returned by an invoked agent is processed before being fed back into the NetGPT core’s reasoning loop. In particular, irrelevant content is masked, and only the informative portion of the response, which is delimited by the tokens \texttt{<ans>} and \texttt{</ans>}, is preserved. This masking clearly separates orchestration logic from response imitation and ensures that policy updates are driven solely by the NetGPT core’s own reasoning and coordination behavior, rather than being influenced by extraneous content or biases from the invoked agents.

\subsubsection{Entropy Monitoring and Exploration Control}
An entropy monitoring process is employed to track token-level uncertainty throughout the reasoning trajectory, especially after interactions with other agents. An increase in entropy indicates rising ambiguity or insufficient internal understanding, which in turn activates dynamic exploration to encourage diversity. Besides, entropy can be incorporated directly into advantage estimation through an uncertainty-aware correction term. Under this mechanism, high-entropy states trigger more exploratory sampling during rollouts, transforming entropy from a static regularization term into an active exploration signal that guides the policy toward informative trajectories, improving effectiveness in long-horizon reasoning.

\subsubsection{Multi-Objective Reward Design}
A multi-objective reward is used to maximize task utility while minimizing Service-Level Agreement (SLA) violations and resource cost, capturing both task-level success and system-level efficiency. 
Accuracy rewards ensure correct and interpretable task completion, while format rewards encourage adherence to the required data structure.
Notably, it is feasible to incorporate additional reward components to support more diversified preferences. For example, efficiency rewards favor shorter reasoning paths and fewer redundant agent invocations; 
quality-of-service rewards penalize excessive latency and computation; 
and exploration rewards promote the discovery of new, high-performing agent combinations. 
Through this multi-objective optimization, NetGPT gradually learns to balance reasoning depth, collaboration cost, and overall network efficiency.

\section{Case Study}
To validate the effectiveness of communication-incentivized collaborative reasoning through agentic RL, we conduct a proof-of-concept experiment by verifying whether the post-training system can correctly interpret intents, invoke appropriate sub-agents, and integrate multi-agent outputs to complete complex network-related tasks. 
\subsection{Experimental Setup}\
We primarily consider the NetGPT core and two predefined agentic actions exposed by NetGPT: 
\begin{itemize}
	\item  Network Analysis: 
	This action type supports the interpretation of network data, performance evaluation, and fault diagnosis. It is implemented through the \texttt{Llama-3.2-3B-Instruct} model fine-tuned on the improved TeleLogs dataset \cite{TelelogsDateset}. 
\item Protocol Query:
This action type enables protocol document search and context retrieval through Retrieval Augmented Generation (RAG) or online search. 
We modify \texttt{Telco-RAG} \cite{TelcoRAG}, combining \texttt{Llama-3.2-3B-Instruct} for response synthesis and \texttt{text-embedding-3-large} for retrieval over a pre-indexed 3GPP protocol corpus containing Technical Specifications (TS) and Technical Reports (TR). 
\end{itemize}

Besides these example action types, the NetGPT core is built on the \texttt{Llama-3-8B-Instruct} model, and directly answers queries related to the TeleQnA dataset \cite{TeleQnAdataset} in a single turn without invoking agents. A collaborative reasoning example of network root cause analysis can be found at the bottom part of Fig. \ref{fig2:operation details}.

\begin{figure*}[t]
	\centering 
    \hspace{-0.2cm}
	\includegraphics[scale = 0.6]{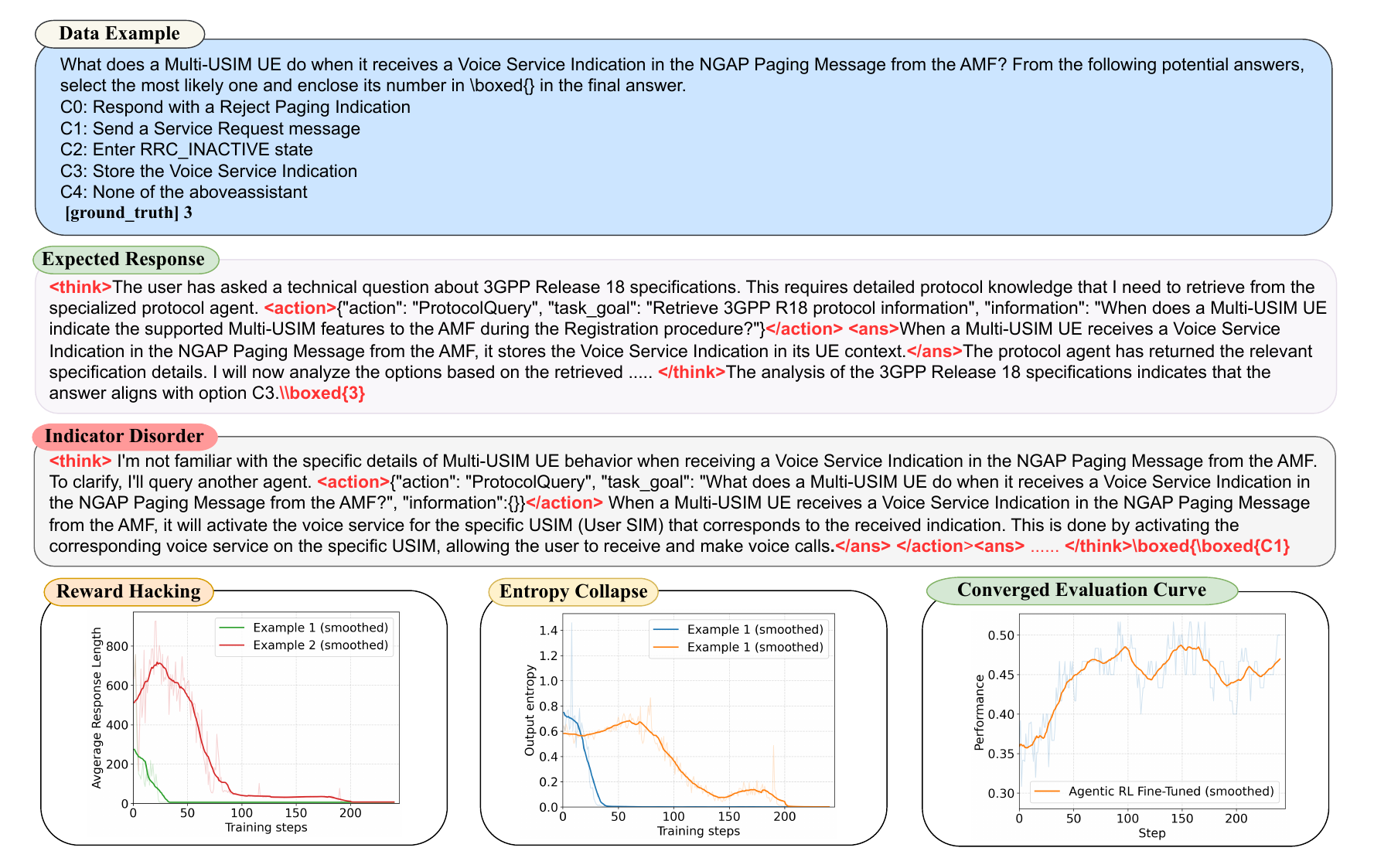}
	\caption{Examples of failure modes and the eventual convergence of agentic RL.}
\label{fig5:badsamples}
\end{figure*} 

\subsection{Analysis}
Fig. \ref{fig5:badsamples} illustrates several typical failure modes that occur when applying prompt-based and conventional RL-based fine-tuning without the careful calibration discussed in Section \ref{sec:training_pipeline}. 
For example, relying solely on prompt-based methods leads to unstable instruction following. As shown in the ``Indicator Disorder'' case, the model confuses key tokens for collaborative reasoning, leading to malformed trajectories and invalid agent invocation. Hence, it necessitates robust supervision for agentic RL. 
Meanwhile, relying solely on scalar and sparse rewards (e.g., RLVR correctness signals) may induce reward hacking, where shortcut behaviors undermine structured reasoning, motivating the need for multi-objective reward design. 
Lastly, insufficient entropy regularization leads to rapid entropy collapse, preventing meaningful exploration. 
Eventually, the concerted efforts in Section \ref{sec:training_pipeline} yield a stable convergent curve in Fig. \ref{fig5:badsamples}. 

\begin{figure}[t]
	\centering 
	\includegraphics[scale = 0.35]{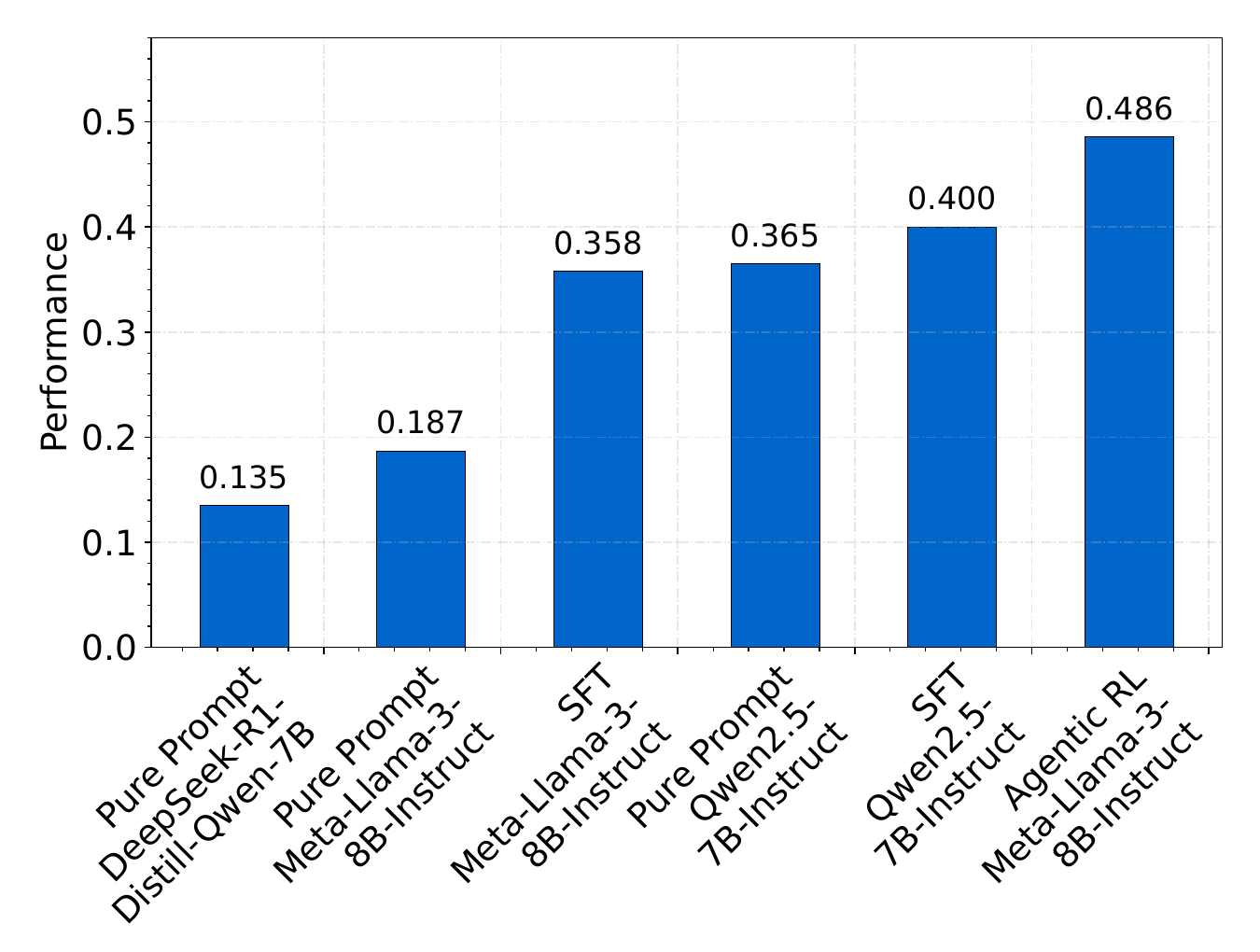}
	\caption{Performance of different methods and models.}
\label{fig6:eval}
\end{figure}

On the other hand, Fig. \ref{fig6:eval} presents the performance comparison between agentic RL-based tuning and other common approaches. It can be observed from Fig. \ref{fig6:eval} that Pure Prompt offers the weakest performance, while SFT provides moderate improvement but remains limited by static supervision. In contrast, agentic RL achieves the highest score, confirming that the collaborative reasoning methodology leads to more effective task accomplishment performance.

\section{Open Research Directions}
While the proposed NetGPT framework demonstrates the feasibility of agentic RL evolvement in xG network intelligence, several open issues can be further investigated. 
\subsection{Reward Definition for Open-Ended Reasoning}\
The reward definition for open-ended reasoning remains a key challenge. Many network-level tasks do not have a standard ground-truth, making it difficult to design consistent reward signals. Future work may consider adaptive or learned reward models that combine expert feedback with outcome-based self-evaluation.
\subsection{Stability and Convergence}
As agents collaborate asynchronously under dynamic network conditions, maintaining smooth policy updates and stable learning will be essential for scaling to more complex and large-scale environments.
\subsection{Scalability and Trustworthiness}
Efficient reasoning compression, lightweight inference, and secure agent coordination are promising directions to enhance the deployability of NetGPT in real xG infrastructures. 

\section{Conclusions}\label{sec:conclusion}
This paper enhances NetGPT to a unified cognitive framework for AI-native xG networks that transforms the communication infrastructure into a powerful collaborative reasoning system. 
By integrating hierarchical cognition, multi-agent collaboration, and continual learning, NetGPT enables autonomous intent understanding, dynamic orchestration, and self-evolving adaptation across the network. 
A two-phase training pipeline has been adopted to realize this capability. 
The SFT stage equips the model with structural and procedural awareness of agent collaboration, while the agentic RL stage refines reasoning depth and coordination. 
The proof-of-concept experiments have verified the feasibility of the framework and the improvement of task completion. 
NetGPT represents an early but tangible step toward self-organizing, cognition-driven xG systems capable of sensing, reasoning, and acting autonomously. 

\bibliographystyle{IEEEtran}
\bibliography{reference}

\section*{Author Biographies}
\textbf{Xiaoxue Yu} (sdwhyxx@zju.edu.cn)  is a PhD Candidate in Zhejiang University, Hangzhou, China. Her research interests currently focus on communications in distributed inference. 

\textbf{Rongpeng Li} (lirongpeng@zju.edu.cn) [corresponding author] is an Associate Professor at Zhejiang University. His research interests currently focus on networked intelligence for comprehensive efficiency (NICE). 

\textbf{Zhifeng Zhao} (zhaozf@zhejianglab.org) is the Chief Engineer with Zhejiang Lab, Hangzhou, China. His research area includes collective intelligence and software-defined networks.

\textbf{Honggang Zhang} (hgzhang@must.edu.mo) is a Professor in Macau University of Science and Technology, Macau, China. He is interested in cognitive green communications.

\end{document}